\documentclass[12pt]{article}
\usepackage{graphicx}
\begin{document}
%
%
\begin{center}
{\large $\pi \rightarrow \pi\pi$ results in nuclei}
\end{center}
%
%
%
%
\begin{center}
\normalsize
P. Camerini$^{a,b}$, E. Fragiacomo$^{a,b}$, N. Grion$^{a,}$\footnote[1]
{Corresponding author, electronic mail: Nevio.Grion@ts.INFN.it}, 
R. Rui$^{a,b}$, J.T. Brack$^{c}, $E.F. Gibson$^{d}$, G.J. Hofman$^{e}$, 
E.L. Mathie$^{f}$, R. Meier$^{g}$, K. Raywood$^e$, M.E. Sevior$^{h}$, 
G.R. Smith$^{e,}$\footnote[2] 
{Present address: Jefferson Lab, Newport News, Va 23006, USA}
and R. Tacik$^{f}$.
\end{center}
%
%

\begin{center}
\small{\it 
$^a$ Istituto Nazionale di Fisica Nucleare, 34127 Trieste, Italy \\
$^b$ Dipartimento di Fisica dell'Universita' di Trieste, 34127 Trieste, 
     Italy\\
$^c$ University of Colorado, Boulder CO 80309-0446, USA \\
$^d$ California State University, Sacramento CA 95819, USA \\
$^e$ TRIUMF, Vancouver, B.C., Canada V6T 2A3 \\
$^f$ University of Regina, Regina, Saskatchewan, Canada S4S 0A2 \\
$^g$ Physikalisches Institut, Universit\"{a}t T\"{u}bingen, 
     72076 T\"{u}bingen, Germany \\
$^h$ School of Physics, University of Melbourne, Parkville, Vic., 3052,
     Australia \\
} 
\end{center}                   

%
%
\setlength{\baselineskip}{2.3ex}         
{\small
The Crystal Ball ($CB$) collaboration at $BNL$ has recently presented
results regarding a study of the $\pi^- A \rightarrow \pi^0\pi^0 A^\prime$ 
reaction on $H, D, C, Al$ and $Cu$, using a nearly 4$\pi$ detector. 
Similar results, but for the $\pi^+ A \rightarrow \pi^+\pi^{\pm} A^\prime$ 
reaction on $^{2}H$, $^{12}C$, $^{40}Ca$, and $^{208}Pb$, have been published 
earlier by the $CHAOS$ collaboration at $TRIUMF$. In this Brief Report a 
comparison of the results of the two measurements is made, which shows that 
the $CHAOS$ and $CB$ data share relevant common features. In particular, the 
increase in strength as a function of A seen in the near-threshold 
$\pi^+ \pi^-$ invariant mass spectra reported by the $CHAOS$ group, is also 
seen in the $\pi^0 \pi^0$ CB data, when the results from the two groups are 
compared in a way which accounts for the different acceptances of the two 
experiments. 
}

\normalsize
%

The phenomenon of nuclear medium modification of $\pi\pi$ properties  
has been the focus of a growing number of recent studies. A selection 
of the most recent theoretical articles can be found in Ref. 
\cite{Selection_theor_articles}. The authors therein have reanalysed 
the modifications caused by nuclear matter at finite density for the 
$\pi\pi$ interaction in the $I=J=0$ channel (the $\sigma$-meson 
channel), and found a dramatic reshaping of the $\pi\pi$ interaction 
strength function at around the $2m_\pi$ threshold. The reshaping
results in a strong enhancement of the strength function around
$2m_\pi$, which is determined by the combined effect of collective 
pionic modes and partial restoration of chiral symmetry. The first 
effect is explained in terms of standard $P-$wave coupling of pions 
to $particle-hole$ and $\Delta-hole$ correlated states. The effect 
of partial restoration of chiral symmetry in nuclear matter relates to
the modification of the basic $(\pi\pi)_{I=J=0}$ interaction: in the 
scalar-isoscalar channel, pion pairs strongly interact with the nucleons
giving rise to the elusive $\sigma-$meson, even at densities below the 
nuclear saturation density. Such a picture of $\pi\pi$ dynamics in nuclear 
matter is supported by the experimental results of the $CHAOS$ collaboration. 
These results, however, seem in part to be contradicted by the recently 
published data from the $CB$ collaboration. The purpose of this Brief Report
is to make a more careful comparison which accounts for the different 
acceptance of the two experiments. Such a comparison reveals that in fact the
main results of the experiments are in good agreement with each other.

The $TRIUMF$ pion production program measured the $\pi^+ A \rightarrow 
\pi^+\pi^{\pm} A^\prime$ reaction on $^{2}H$, $^{12}C$, $^{40}Ca$, and 
$^{208}Pb$ at $p_{\pi^+}=399$ MeV/c, using the $CHAOS$ spectrometer
\cite{CHAOS:spectr}. $CHAOS$ consists of a dipole magnet producing a 
vertical field and four rings 
of cylindrical wire chambers for the tracking of charged particles. The 
outermost wire chamber is surrounded by a segmented telescope for particle 
identification. Particles from a pion production reaction ($\pi, 2\pi$) are 
measured in the horizontal plane (360$^\circ$), and $\pm 7^\circ$ out of this 
plane for a solid angle of $\sim$ 10\% of $4\pi$ sr. The $CB$ collaboration
reported results from a study of the $\pi^- A \rightarrow \pi^0\pi^0 A^\prime$ 
reaction on $H, ^{2}H, C, Al$ and $Cu$ at $p_{\pi^-}=408$ MeV/c, using a 
$NaI(Tl)$ crystal ball which covered 93\% of 4$\pi$ sr \cite{CB:one}. The 
detector is a multiphoton spectrometer, thus  capable of detecting 
$\pi^\circ$'s in the final state. Charged particles are vetoed by a plastic 
scintillator barrel which surrounds the target. Table 1 reports the quantities 
which are relevant to the further discussion.
\begin{table}[bc]
\caption[Table]
{
Relevant quantities in the $CB$ and $CHAOS$ measurements. The notations 
$T_\pi$ and $M_{\pi\pi}$ indicate the single pion kinetic energy and the
$\pi\pi$ invariant mass, respectively.
}
\begin{center}
  \begin{tabular}{cccccc}                                      
\hline 
Detector &  Magnetic field & Solid angle & $T_\pi$ threshold & $M_{\pi\pi}$ resolution &
  Reaction channel   \\ 
         &       [T]       &  [$4\pi$]   &       [MeV]       &        [MeV]            &
    [$\pi\rightarrow\pi\pi$]    \\ 
\hline
   $CB$  &       0.0       &     93\%    &       $\sim$ 0    &      $  3.3 - 7.1  $    &

  $\pi^-\rightarrow\pi^0\pi^0$  \\  
 $CHAOS$ &       0.5       &     10\%    &      $\sim$ 11    &         $ 1.1 -2.1$     &     
 $\pi^+\rightarrow\pi^+\pi^\pm$ \\  
\hline
\end{tabular}
\end{center}
\end{table}
     Since the $CB$ detector has a larger solid angle coverage, the $CB$ data 
     could in principle be analysed and reduced to the $CHAOS$ solid angle. 
     In this manner, the results of the two measurements could be directly 
     compared. This approach, unfortunately, is impractical because of the 
     low statistics of the $CB$ data. In the case of $H$ (Fig. 2 of Ref. 
     \cite{CB:one}), the low-energy yield at $\sim$284 MeV nearly equals the 
     standard deviation thus implying that the bin content is $\sim$1. About 
     the same amount is accumulated in the $\sim$293 MeV bin. As well, the 
     deuterium invariant mass ($M_{\pi\pi}$) spectrum (Fig. 3) displays the 
     same features. The comparison will therefore be addressed to the bulk 
     of the ($\pi, 2\pi$) results.

     Analyses of the $CHAOS$ work focused on studies of the $A$-dependence of the 
     reaction at threshold \cite{CHAOS:one}, the reaction mechanism\cite{CHAOS:two}, 
     the properties of the $\pi\pi$ system in vacuum\cite{CHAOS:three,CHAOS:3.5} and 
     in nuclear matter \cite{CHAOS:four,CHAOS:five}. Some of the previously published 
     $CHAOS$ results most relevant to this Brief Report are summarised in the following. 

    The $\pi \rightarrow \pi\pi$ reaction in nuclei is a quasifree
    process, which involves a single nucleon and proceeds via
    $\pi N \rightarrow \pi\pi N$ \cite{CHAOS:two}.

    Near the $2m_{\pi}$ threshold, the $\pi^+\pi^-$ interacting system 
    predominantly couples in $S$-wave ($\sim$95\%), while $P$-wave coupling 
    is negligible \cite{CHAOS:one,CHAOS:five}.  Together with the arguments
    presented in Ref.\cite{isospin00:one}, this permits the quantum numbers  
    $I=J=0$ to be assigned to the  $\pi^+\pi^-$ system. The same quantum 
    numbers are readily reached by the pure isospin 0 $\pi^0\pi^0$ system 
    in the $CB$ measurement. Therefore, the $\pi^+\pi^-$ system can be 
    directly compared to the $\pi^0\pi^0$ system.

    The most striking physics result from the CHAOS data is the remarkable
    A-dependence observed in the $\pi^+\pi^-$ invariant mass ($M_{\pi\pi}$) 
    distributions at around the $2m_{\pi}$ threshold \cite{CHAOS:five}. 
    Unfortunately the data also exhibit a peak in this same region which 
    is an artifact of the limited out-of-plane acceptance of the spectrometer. 
    The $M_{\pi\pi}$ resolution (1.1-2.1 MeV) does not exceed the $M_{\pi\pi}$
    binning (8.7 MeV), so resolution does not affect the shape of the 
    $M_{\pi\pi}$ distributions. The physics 
    is in the observed A-dependence of the near threshold $M_{\pi\pi}$ 
    distributions, not in the peak itself. Furthermore, the peak is also 
    observed in the simultaneously measured isospin 2 $\pi^+\pi^+$ invariant 
    mass distributions, which exhibit no remarkable A-dependence, and appears 
    in the phase-space simulations. This has been consistently explained in 
    each of the CHAOS articles. 

    The observable $\cal C$$_{\pi\pi}^A$ was defined in Ref.\cite{CHAOS:four}
    precisely in order to disentangle the acceptance issue from 
    the physics observed in the A-dependence, and therefore focus exclusively
    on medium effects.  $\cal C$$_{\pi\pi}^A$ is the composite ratio 
    $\frac{M_{\pi\pi}^A}{\sigma_T^A} / \frac{M_{\pi\pi}^N}{\sigma_T^N}$, 
    where $\sigma_T^A$ ($\sigma_T^N$) is the measured total cross section 
    of the $(\pi, 2\pi)$ process in nuclei (nucleon). This observable was shown 
    to  yield the net effect of nuclear matter on the $\pi\pi$ interacting system, 
    regardless of the reaction mechanism used to produce the pion pair, and to 
    be nearly unrelated to the $CHAOS$ acceptance. These traits and the points 
    discussed above ensure that the composite ratio $\cal C$$_{\pi\pi}^A$ 
    should be nearly the same for the $CHAOS$ and the $CB$ measurements, 
    despite the different solid angles subtended by the two experiments.

    Before using the acceptance independent $\cal C$$_{\pi\pi}^A$ to make a 
    meaningful comparison of the $CB$ \cite{CB:one} and the $CHAOS$ $[4-9]$ 
    results, some comments on the $CB$ results are needed.
    The experimental data for the $2\pi^0$ invariant mass distributions 
    (Figs. 2 and 3 of Ref. \cite{CB:one}) are in general characterised by 
    low statistics. Only a few ($\pi, 2\pi$) events were accumulated in the 
    low-energy yield for $H$ and $^{2}H$. This is very unfortunate, since 
    these poor statistics occur exactly in the most interesting region of 
    the invariant mass distribution, where $\cal C$$_{\pi\pi}^A$ departs 
    from a flat behaviour \cite{CHAOS:four}. 

    Fig. 3 of Ref.\cite{CB:one} shows the experimental $2\pi^0$ invariant mass 
    distributions for the studied nuclei corrected for the $CB$ acceptance. The 
    $CB$ data (in arbitrary units) are compared to the model calculation of 
    Vicente-Vacas \cite{Vicente:one}, dashed line, and Rapp \cite{Rapp:one}, full 
    line. For $H$ (and $^{2}H$, see Ref. \cite{CHAOS:five}), the curves are able 
    to reproduce the shape of the $M_{\pi\pi}$ experimental distribution. In 
    the case of $C$, the widening of the distribution toward high energies reflects 
    the Fermi motion of the struck nucleon, 
    $\pi^- p[A-1] \rightarrow \pi^0\pi^0 n[A-1]$. 
    For $Al$ and $Cu$, these features seem contradicted by the shrinking of the 
    distributions. The $M_{\pi\pi}$ high-energy tail appears depleted rather than 
    further expanded even though the models, which include the Fermi motion, 
    predict a non-negligible yield 20-30 MeV above the $CB$ data.   

    The $2\pi^0$ invariant masses are given in arbitrary units in Ref. \cite{CB:one},
    which precludes comparison to an absolute prediction. Normalizing simulations
    (Fig. 2) and model predictions (Fig. 3) to data can lead to ambiguous 
    interpretations. For instance, the $Cu$-target data in Fig. 3 are poorly 
    described by the curves of Vicente-Vacas (dashed line) and Rapp (full line), 
    thus any analysis is inconclusive. If the curves were normalised to the 
    $M_{\pi\pi}$ high-energy tail, then the data would have explicitly displayed an 
    accumulation of strength at threshold. On the other hand, if the curves were 
    normalized to low-energy invariant masses, then the comparison would have 
    disclosed a rather strong depletion of the $M_{\pi\pi}$ high-energy yield. 
    The case for the hydrogen target is different: the theoretical predictions 
    seem to be normalized to the experimental data. The calculation by Vicente-Vacas 
    agrees almost perfectly with the data, which reflects the detailed microscopic 
    description of the elementary $\pi N \rightarrow \pi\pi N$ reaction. The model 
    of Rapp, which misses some important diagrams, overestimates the $M_{\pi\pi}$ 
    yields (from 2 to 3 times) in the low-energy interval. The same quantitative 
    conclusion for the hydrogen target is also reached using the CHAOS data 
    \cite{CHAOS:five}. 

    Finally, the results of  $CHAOS$  are compared to the results from the $CB$ 
    Collaboration. The authors of Ref. \cite{CB:one} repeatedly make the point 
    that they do not observe the peak in the near threshold invariant mass that
    is observed by CHAOS. As noted above, this is merely an expected consequence 
    of the different acceptances of the two detectors.

     The results shown in Fig. 4 of Ref. \cite{CB:one} lead to the 
     conclusion that {\em the nuclear medium as represented by nuclei D, 
     C, Al and Cu changes the $\pi-\pi$ interaction in such a way that the 
     $\pi^0\pi^0$ invariant mass distribution from complex targets is closer 
     to the phase-space distribution than is the same distribution from H}. 
     These broad features can also be found in the $CHAOS$ data: the 
     phase-space simulations for the $\pi A \rightarrow \pi\pi N[A-1]$ 
     reaction reported in Fig. 9 of Ref. \cite{CHAOS:five} (shaded regions) 
     should just be compared to the $\pi^+\pi^-$ invariant mass distributions
     (diamonds). The higher A $\pi^+\pi^-$ results clearly match the phase 
     space calculations better than the lower A $\pi^+\pi^-$ results do. 
     These findings, however, do not disclose the nature of the threshold 
     enhancement of the $\pi^+\pi^-$ invariant mass distributions. As pointed 
     out by the $CHAOS$ results, comparison of $M_{\pi^+\pi^-}^A$ with 
     $M_{\pi^+\pi^+}^A$   provides useful insight into $\pi\pi$ dynamics in 
     nuclear matter. Fig. 9 of Ref. \cite{CHAOS:five} shows that the 
     $M_{\pi^+\pi^+}^A$ distributions are accounted for well by the 
     $\pi A\rightarrow\pi\pi N[A-1]$ phase-space from $^{2}H$ to $^{12}C$ to 
     $^{40}Ca$ and to $^{208}Pb$, and the maximum invariant mass increases 
     with the increase of the nucleon Fermi momentum (i.e., $A$), as one would 
     expect. In contrast, the $M_{\pi^+\pi^-}^A$ strength near threshold 
     is negligible for $^2H$,  and strongly increases as $A$ increases. This 
     enhancement is attributed to medium modifications in the $(\pi\pi)_{I=J=0}$ 
     interacting system \cite{CHAOS:four}.
   
     In the $I=J=0$ channel, both the $CB$ and $CHAOS$ measurements find the 
     $M_{\pi\pi}^A$ yield to be close to zero near the $2m_\pi$ threshold for 
     $A=$(1 or) 2, while it increases with increasing $A$. The composite ratio 
     $\cal C$$_{\pi^+\pi^-}^A$  discussed in Refs. \cite{CHAOS:four,CHAOS:five} 
     can be derived from the $CB$ data by dividing the $M_{\pi^0\pi^0}^A$ by the
\begin{figure}[t,b]
 \centering
  \includegraphics*[angle=90,width=0.6\textwidth]{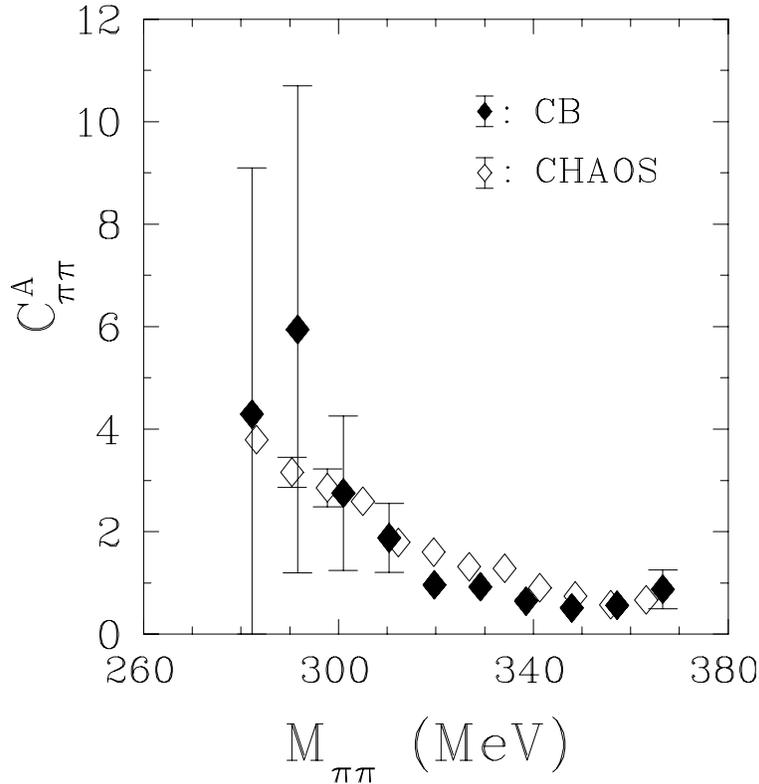}
  \setlength{\abovecaptionskip}{7pt}  
  \setlength{\belowcaptionskip}{0pt}  
  \caption{\footnotesize The composite ratios $\cal C$$_{\pi\pi}^A$ as a 
      function of the $\pi\pi$ invariant mass for the $^{12}C$ target. Full 
      diamonds, the $\cal C$$_{\pi^0\pi^0}^C$  distribution  deduced from 
      the data of Ref. \cite{CB:one} as explained in the text; open diamonds, 
      the $CHAOS$ $\cal C$$_{\pi^+\pi^-}^C$ distribution  taken from 
      \cite{CHAOS:five}.}
\end{figure}
     $M_{\pi^0\pi^0}^H$ yields. $\cal C$$_{\pi^0\pi^0}^A$ for the $C$-target 
     (the only nucleus in common to the two experiments) is shown in Fig. 1 
     with full diamonds.  As discussed above, $\cal C$$_{\pi^0\pi^0}^C$ can
     be directly  compared with $\cal C$$_{\pi^+\pi^-}^C$. Since the 
     $M_{\pi^0\pi^0}^A$ distributions were given in arbitrary units, 
     $\cal C$$_{\pi^0\pi^0}^C$ is normalized to $\cal C$$_{\pi^+\pi^-}^C$ in the 
     energy range above 350 MeV, where the two distributions are flat.
     The large error bars associated with $\cal C$$_{\pi^0\pi^0}^C$ reflect 
     the poor statistical content of $M_{\pi^0\pi^0}^H$ around threshold. 
     It is worthwhile comparing the $CB$ and $CHAOS$ ratios 
     for $C$ since they convey relevant common features: they both are flat at 
     $\sim$350 MeV, are peaked at threshold and are comparable in
     strength. In fact the agreement between the two experiments is
     remarkably good.
     The curious depletion of the invariant mass yield for $Al$ and $Cu$ 
     observed in Ref. \cite{CB:one} makes a close comparison between 
     $\cal C$s' for these two targets of questionable significance.
     Their $\cal C$$_{\pi^0\pi^0}^A$ 
     distributions, however,  resemble the $\cal C$$_{\pi^0\pi^0}^C$ one. 

In summary, the number of current international workshops \cite{workshops:one}
devoted to the study of hadronic properties in nuclear matter reveals the 
increasing importance of this field of nuclear physics. In this realm, the 
correlated system of two pions in the $I=J=0$ channel embodies a special role: 
$(\pi\pi)_{I=J=0}$ is the lighter object carrying the quantum numbers of the 
$QCD$ vacuum. On the other hand, experimental studies of the behaviour of pion 
pairs in nuclear matter are fairly recent. The $CB$ measurement examined the 
$A$-dependence of pion-induced pion production at 408 MeV/c in the 
$\pi^0\pi^0$ neutral channel. An earlier study undertaken by the $CHAOS$ 
collaboration  reported on the $\pi^+\pi^{\pm}$ channels at an incident pion 
momentum of 399 MeV/c. As far as the $(\pi\pi)_{I=J=0}$ interacting system is 
concerned, the $CB$ results share relevant common features with the $CHAOS$ 
results.  When the two experiments are compared using an observable designed 
to mitigate acceptance differences, the results agree with each other very well, 
although the statistical quality of the $CB$ data in the most interesting, 
near-threshold region is marginal.  This agreement may be interpreted as
an indipendent confirmation by the $CB$ measurement of the nuclear matter
modifications in the $I=J=0$ $\pi\pi$ system, first reported by $CHAOS$.

%
%
 
%
%

\begin{thebibliography}{99}
%
        %
        %
\bibitem{Selection_theor_articles}
D. Davesne. Y. J. Zhang and G. Chanfray, Phys. Rev. {\bf C62}, 024604 (2000);
Z. Aouissat, G. Chanfray, P. Schuck, and J. Wambach, Phys. Rev. {\bf C61}, 
012202 (2000); D. Jido, T. Hatsuda and T. Kunihiro, Phys. Rev. {\bf D63}, 
011901-1 (2000)D; T. Hatsuda, T. Kunihiro and H. Shimizu,  Phys. Rev. Lett.
{\bf 63}, 2840 (1999).  
        %
        %
             \bibitem{CHAOS:spectr} 
             G.R. Smith {\em et al.}, Nucl. Instr. and Meth. in Phys. 
             Res. {\bf A362}, 349 (1995). Further details on $CHAOS$ 
             can be found in the references therein quoted.
        %
        %
\bibitem{CB:one}
A. Starostin {\em et al.}, Phys. Rev. Lett. {\bf 85}, 5539 (2000).
        %
        %
    \bibitem{CHAOS:one}
      F. Bonutti {\em et al.}, Phys. Rev. Lett. {\bf 77}, 603 (1996).
    %
    \bibitem{CHAOS:two}
      F. Bonutti {\em et al.}, Phys. Rev. {\bf C55}, 2998 (1997).
    %
    \bibitem{CHAOS:three}
    F. Bonutti {\em et al.}, Nucl. Phys. {\bf A638}, 729 (1998).
    %
    \bibitem{CHAOS:3.5}
    M. Kermani {\em et al.}, Phys. Rev. {\bf C58}, 3419 (1998).
    %
    \bibitem{CHAOS:four}
      F. Bonutti {\em et al.}, Phys. Rev. {\bf C60}, 018201 (1999).
    %
    \bibitem{CHAOS:five}
      F. Bonutti {\em et al.}, Nucl. Phys. {\bf A677}, 213 (2000).
%
%
\bibitem{isospin00:one}
D. Lohse, J.W. Durso, K. Olinde and J. Speth, Phys. Lett. {\bf B234}, 
235 (1990).
%
        %
        %
\bibitem{Vicente:one}
  M. J. Vicente-Vacas and E. Oset, Phys. Rev. {\bf C60}, 064621 (1999).
%
\bibitem{Rapp:one}
        R. Rapp, J. W. Durso, Z. Aouissat G. Chanfray, O. Krehl, P. Schuck 
        J. Speth and J. Wambach, Phys. Rev. {\bf C59}, R1237 (1999).
%
%
\bibitem{workshops:one}
Proceedings of the international workshops on
{\it $\sigma$-Meson and Hadron Physics}, KEK Proceedings 2000-4;
and {\it Workshop XXVIII on Gross Properties of Nuclei and 
Nuclear Excitations}, Hirschegg (Austria); 
and workshop on {\it Chiral Fluctuations in Hadronic Matter}, September 
26-28 2001, Paris (France).   
%
\end{thebibliography}
\end{document}